\documentclass[12pt]{article}
\usepackage{cite}
\usepackage{amsmath,amsfonts,amssymb}
\usepackage{graphicx,color}
\usepackage{epsfig,epsf}
\usepackage{bm}

\usepackage{relsize}

\usepackage[bookmarks, breaklinks, colorlinks,urlcolor=black, citecolor=red, 
linkcolor=blue]{hyperref}

\usepackage{url}
\usepackage{multirow}
\usepackage{slashed}

\usepackage{ulem}

\def\bea{\begin{eqnarray}}
\def\eea{\end{eqnarray}} 
\def\be{\begin{equation}}
\def\ee{\end{equation}} 
\def\nn {\nonumber}

\def \Im{\text{Im}}
\def\gev{\ensuremath{\mathrm{Ge\kern -0.1em V}}}
\def\mev{\ensuremath{\mathrm{Me\kern -0.1em V}}}

\def\w0q{\ensuremath{\omega_0^{(q)}}}

\definecolor{Darkgreen}{RGB}{30,150,30}

\begin{document}

\begin{flushright}
SI-HEP-2020-20\\
P3H-20-039 \\[0.2cm]
\end{flushright}


\vskip 2cm	

\begin{center}
	
	{\Large\bf 
	Inverse moment of the $B_s$-meson
        distribution \\[3mm] amplitude  from QCD sum rule}
	\\[10mm]
	{Alexander Khodjamirian\,\footnote{Email: khodjamirian@physik.uni-siegen.de}, 
Rusa Mandal\,\footnote{Email: Rusa.Mandal@uni-siegen.de}
		and  Thomas Mannel\,\footnote{Email: mannel@physik.uni-siegen.de }
}
\\[6pt]
	
	{\small\it Theoretische Physik 1,
          Naturwissenschaftlich-Technische Fakult\"at, \\ Universit\"at Siegen, 57068 Siegen, Germany}

\end{center}


\begin{abstract}

\noindent 
We derive a QCD sum rule for the inverse moment of the $B_s$-meson light-cone distribution amplitude in HQET. Within this method, the $SU(3)_{fl}$ symmetry violation is traced to the strange quark mass and to the  difference between strange and nonstrange quark condensate densities. We predict the ratio of  inverse moments $\lambda_{B_s}/\lambda_B= 1.19 \pm 0.14$ which can be used in various applications of these distribution amplitudes to the analyses of $B_{s}$-meson decays, provided an accurate value of $\lambda_B$ is available from other sources, such as the $B\to \ell \nu_\ell \gamma$ decay.

\end{abstract}

\section{Introduction}
\label{sec:Intro}

Heavy mesons carrying bottom and strange quantum numbers have
attracted increasing attention due to the large data sample
which has been collected by the LHCb collaboration over the last
years. With the new precise
measurements it becomes increasingly important to also get
more accurate theoretical predictions. This
involves, in particular, reliable estimates of the $SU(3)$
flavour symmetry violation, to be taken into account while analyzing  the $B_s$
data and comparing to the results for nonstrange bottom mesons.
Among important quantities in the theoretical analyses are the light-cone
distribution amplitudes (DAs) of heavy mesons. 
While these quantities were extensively studied for nonstrange bottom mesons,
the $SU(3)_{fl}$ violating effects for $B_s$ have not yet been estimated.   

The light-cone DAs of $B$-meson introduced \cite{Grozin:1996pq}
in the framework of Heavy Quark Effective Theory (HQET) (see also \cite{Szczepaniak:1990dt})
describe  momentum distribution of the light quark in a heavy
pseudoscalar meson.
These DAs enter various factorization formulas
for the exclusive decays of $B$-meson (see e.g. \cite{Beneke:2000wa,Beneke:1999br,Korchemsky:1999qb,DescotesGenon:2002mw,Bosch:2003fc}). They also provide 
the main nonperturbative input in one of the versions \cite{Khodjamirian:2005ea}
of QCD light-cone sum rules for $B$-meson  form
factors. In all these applications, the key parameter 
is the inverse moment of the leading (lowest twist) $B$-meson DA.  

Evidently, the mass of the light $u,d$ quarks plays no role
in the $B_{u,d}$ meson DAs. It can safely
be neglected not only in comparison with any
of the large scales involved in an exclusive $B_{u,d}$ decay,
but also  with respect to typical hadronic scales of $\mathcal{O}
(\Lambda_{QCD})$. This is, however, not the case for the
  $s$-quark. In fact, 
the leptonic decay constants of bottom mesons  
exhibit an appreciable $SU(3)_{fl}$ symmetry  violation.  
The ratio $f_{B_s}/f_{B_{u,d}}$ calculated from the lattice QCD 
deviates by about $20\%$ from the unity \cite{Aoki:2019cca}.  
QCD sum rules (see e.g. \cite{Jamin:2001fw,Gelhausen:2013wia}) 
predict  this ratio in the same ballpark. 
Nevertheless, the influence of  the strange quark
mass on the inverse moment of $B_s$-meson DA has never been investigated.
One of the reasons is that the heavy meson DAs are not yet accessible in 
lattice QCD (for the first exploratory studies see e.g. \cite{Kane:2019jtj,Desiderio:2020oej}). For simplicity,  the inverse moments of all bottom
mesons are assumed equal, as for example in the QCD factorization  analysis of 
nonleptonic $B$ and $B_{s}$ decays \cite{Beneke:2003zv}. 

The $B_s$-meson DA is needed  to describe many important 
decay channels, 
such as the $\bar{B}_s\to  K^{(*)},\phi $ semileptonic and
Flavor Changing Neutral Current (FCNC) transitions, as well as 
various nonleptonic $B_s$ decays, where precision predictions for observables 
are vitally needed. It is therefore timely to make a quantitative
assessment  of the $SU(3)_{fl}$ symmetry violation in  
the bottom meson  DAs. 

In the  future, accurate measurements of the photoleptonic decay 
$B^-\to \ell^-\bar{\nu}_\ell \gamma$ will allow to constrain 
the inverse moment of the $B$-meson DA  using a well elaborated factorization 
formula for  the form factors 
of this decay (the most recent analyses can be 
found in \cite{Beneke:2011nf,Beneke:2018wjp,Wang:2018wfj}). 
There is no such channel available for the $B_s$-meson DA. 
For example, the FCNC decay $B_s\to \ell^+\ell^-\gamma$ is 
``contaminated''  by nonlocal hadronic effects which are not simply 
reducible to DAs. In this situation, a theory estimate of the $SU(3)_{fl}$ 
violation in the  inverse moment is definitely useful.

In this paper,  we obtain the inverse moments of the $B_s$  meson from  the 
QCD sum rule in HQET.
We closely follow the method  used in \cite{Grozin:1996pq,Braun:2003wx} for
the $B$\,- meson DA, but, in contrast, we do not attempt
to determine the shape of DA. 
Instead,  we obtain a QCD sum rule 
directly for the inverse moment. Including  the $\mathcal{O}(m_s)$ effects in the perturbative part and taking into account the difference between strange and nonstrange quark condensates,
we estimate
the inverse moment of $B_s$-meson DA and, as a byproduct, the inverse
moment of the $B$-meson DA. Finally, we predict the ratio of the two
inverse moments with  a lesser uncertainty.

 In what follows, in Section~\ref{sect:SR} we specify the method 
combining  the two  QCD sum rules in HQET: the one 
for the leptonic decay constant and the another one for the DA.
The effects of strange quark mass are calculated and taken into
account. Our numerical results are presented in Section~\ref{sect:Num} and we summarize in Section~\ref{sect:Sum}. The Appendix contains
some useful details of the calculation.

\section{The method}
\label{sect:SR}

\subsection{Sum rule for the $B_s$ decay constant}
\label{sec:decayconst}

To explain how the $SU(3)_{fl}$ violating difference 
between strange and nonstrange bottom mesons
emerges, it is instructive to begin with
the  sum rule for the leptonic decay
constant. We consider it first in QCD with finite masses of $b$
 and $s$ quarks, and  then take the limit of infinitely heavy
 $b$-quark, performing a transition to the HQET sum rule. 
We also need the latter sum rule to fix
the input parameters for the sum rule determination of the $B_{s}$-meson DAs.

We start with the  correlation function of the 
two pseudoscalar heavy-light currents defined in the standard way:
\bea
\label{eq:corrFb}
\Pi_{5q}(q^2) = i \!\int \!d^4 x\, e^{iq\cdot x} \langle 0|
T \{ j_{5q}(x) j_{5q}^\dagger(0) \} | 0\rangle\,,
\eea
where $j_{5q}$ is the divergence of the axial current 
$j_{5q} = (m_b + m_q)\bar{q}i \gamma_5 b$ 
with $m_b$ and $m_q$ being the $b$-quark
and  light-quark ($q=u,d,s$) mass, respectively.
In what follows, we neglect the $u,d$-quark masses,  assuming isospin and chiral symmetry. We also  adopt the
$\overline{MS}$ scheme for the $s$- and $b$-quark masses.
The decay constant of the pseudoscalar $B \equiv
B_{u,d}$- and $B_s$-meson is defined, respectively, as
\bea
\label{eq:fB}
\langle 0| j_{5u(d)} | B (p_B)\rangle = m_{B}^2f_{B}\,,~~
\langle 0| j_{5s} | B_s(p_{B_s})\rangle = m_{B_s}^2f_{B_s}\,.
\eea
The correlation function \eqref{eq:corrFb} satisfies a
double-subtracted dispersion relation which, after the Borel transformation
$|q^2|\to M^2$, takes the form 
\bea
\label{eq:disp}
\Pi_{5q}(M^2) = \frac{1}{\pi}
\int\limits_{0}^{\infty}\! ds\, e^{-s/M^2}\mbox{Im}\Pi_{5q}(s)\,,
\eea
 in which subtraction terms vanish. 
To obtain the sum rule, we use the operator product expansion (OPE) of 
the correlation function $\Pi_{5q}$ valid at deep
spacelike $q^2\ll m_b^2$ or, equivalently, at sufficiently large $M^2$.
The   result $\Pi_{5q}^{\rm (OPE)}(M^2)$ consists of the  perturbative and nonperturbative
(vacuum condensate) parts:
\bea
\label{eq:PiOPE}
\Pi^{\rm(OPE)}_{5q}(M^2)=\int\limits_{(m_b+m_s)^2}^{\infty}\! ds\,
e^{-s/M^2}\rho^{\rm (pert)}_{5q}(s)+\Pi^{\rm(cond)}_{5q}(M^2) \,,
\eea
where the perturbative
contribution is written 
in a dispersion integral form with the  spectral density
$$\rho^{\rm (pert)}_{5q}(s)= \frac{1}{\pi}\Im\, \Pi^{\rm (pert)}_{5q}(s)\,.$$
Adopting the usual quark-hadron duality
 ansatz, the hadronic spectral density in
  Eq.~(\ref{eq:disp}) is approximated by the contribution 
of the lowest  pseudoscalar bottom meson and the
OPE perturbative density taken above an effective threshold.
Considering, for definiteness, the $B_s$ case of our interest,
we have:
\bea
\label{eq:Impi}
\frac{1}{\pi}\mbox{Im}\Pi_{5s}(s) = m_{B_s}^4 f_{B_s}^2 \delta (s -
m_{B_s}^2) 
+ \theta(s -s_{0s}) 
\rho^{\rm (pert)}_{5s}(s)\,.
\eea
Substituting the above expression in r.h.s. of
  Eq.~(\ref{eq:disp})  and using for the l.h.s. the OPE result  
(\ref{eq:PiOPE}), we arrive at  the sum rule:
\bea
\label{eq:SRfB}
m_{B_s}^4 f_{B_s}^2 e^{-m_{B_s}^2/M^2} = \!\! \!\!
\int\limits_{(m_b+m_s)^2}^{s_{0s}} \!\!\!ds \, \rho_{5s}^{\rm(pert)}(s)
e^{-s/M^2}  + \Pi^{\rm(cond)}_{5q}(M^2)\,.
\eea
The leading-order (LO) perturbative term in the OPE of the spectral density
arises from the simple quark-antiquark loop diagram and is  given by~
\bea
\rho^{\rm (pert,LO)}_{5s}(s)= \frac{3}{8\pi^2}(m_b+m_s)^2  \left(1-
 \frac{(m_b-m_s)^2}{s} \right)\lambda^{1/2}(s,m_b^2,m_s^2)\,,
\label{eq:LOrho}
\eea
where $\lambda(x,y,z)\equiv x^2+y^2+z^2-2xy-2xz-2yz$ is the K\"all\'en
function. 
Note that here it is more convenient to use the above spectral density
than to expand  it in the powers of $m_s$ as it is customary in the literature 
(see e.g., \cite{Jamin:2001fw,Gelhausen:2013wia}).
To complete the sum rule, the gluon radiative 
corrections will  be added to the r.h.s. of Eq.~(\ref{eq:SRfB}). 
All necessary expressions
can be found  e.g., in \cite{Gelhausen:2013wia} .

As a next step,  we transform the variables and parameters in the sum
rule (\ref{eq:SRfB}) in order to separate the 
heavy $b$-quark scale and pave the way to the
sum rule in HQET. In our case, there is a nonvanishing 
$s$-quark mass involved  in this transformation.
We
express the external momentum squared in the correlation 
function (\ref{eq:corrFb}) in terms of a new variable $\omega$:
\be
q^2=m_b^2+2m_b\,\omega\,,
\label{eq:q2om}
\ee
and, simultaneously, replace the $B_s$-meson mass by 
\be
m_{B_s}= m_b + \bar{\Lambda}_s\,,
\label{eq:mBs}
\ee 
so that $\omega$ and $\bar{\Lambda}_s$ do 
not scale with the $b$-quark  mass.

The relation (\ref{eq:q2om}) yields for the variable $q^2=s$ in the timelike region: 
\bea
\label{eq:someg}
s=m_b^2+2m_b\omega'\,, 
\eea 
so that $\omega'$ will serve as the integration variable in the sum rule.
According to Eq.~(\ref{eq:mBs}),
the position of the $B_s$ pole at $s=m_{B_s}^2$ corresponds 
to $\omega'= \bar{\Lambda}_s$ and the quark-antiquark threshold of the 
loop diagram at $s=(m_b+m_s)^2$ turns into
$\omega'=m_s$. Note that all these relations are valid  up to
$O(1/m_b)$  corrections which vanish in the $m_b\to \infty$ limit. 
Furthermore, in accordance with the above definitions, we  transform the threshold and the  Borel parameter, respectively:
\bea
\label{eq:hqet2}
s_{0s}= m_b^2 + 2m_b\,\omega_{0s},~~~ \mbox{and} ~~~M^2= 2 m_b \tau\,,
\eea
where the parameter $\omega_{0s}$ and
the variable $\tau$  again do not scale with the $b$-quark mass.
Note, on the other hand, that  the $B_s$ binding energy
$\bar{\Lambda}_s$ and the effective threshold $\omega_{0s}$ 
 both implicitly depend on $m_s$.
Applying Eqs.~(\ref{eq:q2om})--(\ref{eq:hqet2}) to  
 Eq.~(\ref{eq:SRfB}), we then take  the limit  $m_b\to \infty$, 
 transforming this sum rule to its HQET form:
\bea
\label{eq:SRF}
\big[F_{B_s}(\mu)\big]^2 e^{-\bar{\Lambda}_s/\tau} &=& \frac{3}{\pi^2} 
\int\limits_{m_s}^{\omega_{0s}} \! d\omega^\prime \,  e^{-\omega^\prime/\tau}  (\omega^\prime +m_s) \sqrt{\omega^{\prime2} -m_s^2} \nn \\
&+&
\frac{3 \alpha_s}{\pi^3}\int\limits_{0}^{\omega_{0s}} \! d\omega^\prime\,  e^{-\omega^\prime/\tau}   \omega^{\prime2}\,
\left( \frac{17}{3}+\frac{4\pi^2}{9}-2\ln\frac{2\omega^\prime}{\mu}\right) 
\nn \\
&-&\langle\bar s s\rangle
\left[1+\frac{2\alpha_s}{\pi}-\frac{m_0^2}{16\tau^2}\right]\,, 
\eea
where $F_{B_s}$ is the decay constant in HQET, related at
 the $\mathcal{O}(\alpha_s)$
accuracy to the one defined in Eq.~(\ref{eq:fB}):  
\bea
f_{B_s}\sqrt{m_{B_s}}= F_{B_s}(\mu)\left[1+\frac{C_F\alpha_s}{4\pi}
\left(3\ln\frac{m_b}{\mu}-2\right)+\ldots\right],
\eea
and $\mu$ is the renormalization scale. The expression for the perturbative LO part coincides with the one in \cite{Broadhurst:1991fc} where the light-quark mass was retained in the heavy-quark limit of the correlation function \eqref{eq:corrFb}.
Furthermore, in the second line of Eq.~(\ref{eq:SRF}) we have added the 
$\mathcal{O}(\alpha_s)$ gluon radiative corrections 
in HQET taking them from \cite{Braun:2003wx}\,\footnote{The relation of these 
corrections 
to the ones \cite{Jamin:2001fw, Gelhausen:2013wia} in the full QCD sum rule deserves a separate discussion for which we refer to  \cite{Bagan:1991sg,Neubert:1991sp,Broadhurst:1991fc}. }.
In the third line we include the
condensate contributions, where $\langle\bar s s\rangle$ 
denotes the strange quark condensate density and $m_0^2$ is the ratio of the
quark-gluon and quark condensates. 
Note that in the sum rule \eqref{eq:SRF}, we
have neglected the very small effects
of $\mathcal{O}(\alpha_sm_s)$ in the perturbative spectral density 
(hence, the zero limit in the second integral in
  Eq.~(\ref{eq:SRF})) as
well as in the quark condensate term. 
In addition, we assume that 
the ratio $m_0^2$ is the same for all three light quarks and 
neglect the 
numerically insignificant contributions of gluon and four-quark condensates.

The sum rule (\ref{eq:SRF}) can also be  
derived in the framework of HQET as it was done for the 
nonstrange $B$-meson in 
\cite{Shuryak:1981fza, Bagan:1991sg, Neubert:1991sp}.
One starts from the correlation function
of currents containing the effective heavy quark field $h_v$,
so that the external four-momentum is $k=q-m_bv$. 
In this case, the effective variable $\omega=k\cdot v$, where 
$v=(1,\vec{0})$ is the velocity four-vector, replaces
$q^2$,  and the deep spacelike region $q^2\ll m_b^2$ corresponds
to the external off-shell energy $\omega\ll 0$. 
Accordingly, the dispersion relation in the variable 
$\omega$ is used with the $B_s$ pole located at
$\omega=\bar{\Lambda}_s$ and the duality interval  $m_s<\omega<\omega_{0s}$.
The HQET state of a $B_{(s)}$-meson differs from the state in  Eq.~(\ref{eq:fB}) by a normalization factor:
\bea
|B_{(s)}(v)\rangle = \big(m_{B_{(s)}}\big)^{-1/2}|B_{(s)}(p_{B_s})\rangle\,.
\label{eq:Bvstate}
\eea

The HQET sum rule for the nonstrange $B$-meson decay constant in the adopted approximation
is simply obtained from Eq.~\eqref{eq:SRF}
putting $m_s\to 0$ and
replacing 
\bea
\label{eq:replacement}
\bar{\Lambda}_s\to \bar{\Lambda}, ~~\omega_{0s}\to \omega_0\,,~~~ 
\langle \bar{s}s\rangle\to \langle\bar{u}u \rangle\simeq
\langle\bar{d}d \rangle\,.
\eea
The definitions (\ref{eq:q2om}) and (\ref{eq:someg}) are
  the same, but the lower limit of the integration over $\omega'$ shifts from $m_s$ to  zero.
A comparison of the sum rules for $F_{B_s}$ and 
$F_B$ reveals several contributions to the 
$SU(3)_{fl}$ violation. One of them is due to the $m_s$-dependence
of the perturbative spectral density in \eqref{eq:SRF}. 
Note that this effect
is of $\mathcal{O} (m_s/\omega_{0})\sim \mathcal{O}(m_s/\tau)$, that is, parametrically enhanced with respect to the terms proportional to $m_s/m_b$ which have vanished in the 
infinitely heavy quark limit.
An additional $SU(3)_{fl}$
violation effect in the OPE of the correlation function revealed
on r.h.s. of Eq.~(\ref{eq:SRF})
is caused by  the difference between the strange and nonstrange 
quark condensate densities. 
This nonperturbative effect intrinsically
depends on the $s$-quark mass, however
this dependence cannot be represented in an explicit 
form.  
In the sum rule,  the $s$-quark mass effects in the OPE are balanced in the hadronic part  by the differences 
between the effective thresholds ($\omega_{0s}$ versus $\omega_{0}$) and 
the binding energies. For the latter we have the   relation:
\bea
\label{eq:lambdarel}
\bar\Lambda_s= \bar{\Lambda} + m_{B_s}- m_B\,.
\eea    

In the numerical analysis below, we will use the sum rule
  \eqref{eq:SRF} and its counterpart for
$B$-meson to estimate the value  of the continuum thresholds $\omega_0$ and $\omega_{0s}$. To this end, each sum rule
is differentiated with respect to $-1/\tau$ and then divided by itself.
In the resulting relations, 
the dependence on the decay constants $F_B$ and $F_{B_s}$ drops
out, allowing us  to fix  $\omega_0$ 
and $\omega_{0s}$ at a certain adopted
value of $\bar{\Lambda}$,
whereas $\bar\Lambda_s$ is given by the
relation (\ref{eq:lambdarel}).

Having revealed the scale of $SU(3)_{fl}$  symmetry violation in the decay constants of heavy-light mesons, we anticipate the effects to be in the same ballpark in more involved hadronic matrix elements, such as the DAs of $B_s$-meson.

\subsection{Sum rule for the inverse moment of $B_s$ DA}

The $B_s$-meson light-cone DA is defined as the hadronic
matrix element of the  bilocal operator built of an effective heavy-quark field $h_v$ with velocity $v$ and a  strange antiquark field $\bar{s}$ located at a lightlike separation:
\bea
\label{eq:DAdef}
\langle 0|\bar s(tn) i\gamma_5\!\not\!n
[tn,0] h_v(0)|\bar {B}_{s}(v)\rangle =
F_{B_s}(\mu)\!\!\int\limits_0^\infty \! dk\, e^{-i k t}\phi^{B_s}_+(k,\mu)\,,
\eea
with the lightlike gauge link
\bea
{}[tn,0] \equiv {\rm P\,exp}\left[ig\int\limits_0^1\!du\,n_\mu A^\mu(utn)\right].
\eea
Here $n_\mu$ is the lightlike vector, $n^2=0$, such that $n\cdot v=1$, and
$t$ is an arbitrary real valued parameter. In
Eq.~(\ref{eq:DAdef}) we used the general definition \cite{Grozin:1996pq,Beneke:2000wa} of the two-particle heavy-meson DA (see e.g. eq. (17) in 
\cite{Khodjamirian:2005ea}) and projected it onto the
leading, twist-2 DA component $\phi^{B_s}_+(\omega)$,
multiplying both sides of this definition by $(i\gamma_5\!\!\not\! n) $ and
taking the trace. Note that $|\bar B_s(v)\rangle$ is the HQET state
defined in Eq.~(\ref{eq:Bvstate}).
The variable $k$ in (\ref{eq:DAdef}) can be interpreted as the
  light-cone projection of the light $s$-quark momentum. 
Due to non-vanishing $m_s$, it is natural to expect that for $B_s$ this
variable is limited from below by $k=m_s$, hence in a realistic 
model, $\phi^{B_s}_+(k,\mu)\sim\theta(k-m_s)$. However, here we will
not dwell on reproducing the shape of the $B_s$-meson DA. 
Instead, we concentrate on our main task, that is, to 
obtain a sum rule estimate for the inverse moment defined as:
\bea
\lambda_{B_{(s)}}^{-1}(\mu) = \int\limits_0^\infty
\frac{dk}{k}\,\phi^{B_{(s)}}_+(k,\mu)\,.
\label{eq:lambBs}
\eea

To achieve the goal, we  largely follow
the method used for the $B$-meson in \cite{Grozin:1996pq} and
 upgraded in \cite{Braun:2003wx} to include the gluon radiative
 corrections (see also the review~\cite{Grozin:2005iz}). At the same time, we modify this method and obtain 
the sum rule directly for the inverse moment (\ref{eq:lambBs}).

To this end, we introduce the following correlation function in HQET:
\bea
\label{eq:BDAsrcorr}
{\cal P}_{s}(\omega,t)=i\!\int\! d^4xe^{-i\omega\, v\cdot x}
\langle 0|T\big\{\bar s(tn)i\gamma_5\!\not\!n [tn,0]  h_v(0) 
\bar{h}_v(x) i\gamma_5s(x)\big\}|0\rangle\,.
\eea
It contains  a product of the bilocal operator 
(\ref{eq:DAdef})  with the  local pseudoscalar current 
interpolating the $B_s(v)$ state. 
The variable $\omega$ is analogous to the one
introduced in  Eq.~(\ref{eq:q2om}) for a transition to HQET
of a simpler two-point correlation function (\ref{eq:corrFb}). 
In other words, if, instead of Eq.~(\ref{eq:BDAsrcorr}), we consider a correlation function with the finite mass $b$-quark fields 
and the external four-momentum $q$, then, after reparameterizing 
to the effective fields $h_v$,
the four-momentum in the exponent becomes $(q-m_bv)=\omega v$. 

The hadronic dispersion relation for the correlation function (\ref{eq:BDAsrcorr})
follows from analyticity
with respect to the effective variable $\omega$:
\bea
\label{eq:BDASdisp}
&&{\cal P}_{s}(\omega,t)=
\frac1{\pi}\int\limits_{0}^\infty 
d\omega'\frac{\mbox{Im}{\cal P}_{s}(\omega,t)}{\omega'-\omega}
\nonumber\\
&&=
\frac{\langle 0|\bar s(tn) i\gamma_5\not\!n [tn,0] h_v(0) |\bar {B}_s(v)\rangle
\langle \bar {B}_s(v) |\bar{h}_v i\gamma_5s|0\rangle}{2(\bar{\Lambda}_s-\omega)}+\dots
\nonumber\\
&&=\frac{\big[\,F_{B_s}(\mu)]^2}{2(\bar{\Lambda}_s-\omega)}
\int\limits_0^\infty \! d k\, e^{-i k t}\phi^{B_s}_+(k,\mu)+\dots
\,,
\eea
In the above, the contribution of the $B_s$ pole located at
$\omega=\bar{\Lambda}_s$ is singled out, and  we use the definition
(\ref{eq:DAdef}) together with the one for the $B_s$ decay constant in HQET:
\bea
\label{eq:DAFBs}
\langle \bar {B}_{s}(v)|\bar{h}_v i\gamma_5 s |0\rangle =F_{B_s}\,.
\eea
As it is usually done in QCD sum rules, the contributions of excited and
continuum states with the $B_s$ quantum numbers, indicated 
in Eq.~(\ref{eq:BDASdisp}) by the ellipsis,
will be approximated assuming the quark-hadron
duality. Furthermore, to decrease the sensitivity to this
approximation, 
we employ the Borel transformation in the variable $\omega$ defined in HQET as:
\be
{\cal B}_\tau f(\omega)=
\lim\limits_{\{-\omega, n\}\to \infty, -\omega/n=\tau}
\frac{(-\omega)^{n+1} }{n!}\left(\frac{d}{d\omega}\right)^n  f(\omega)
\equiv f(\tau)\,.
\label{eq:Boreltau}
\ee
Applying it to Eq.~(\ref{eq:BDASdisp}), we have:
\bea
\label{eq:BDASdisp2}
{\cal P}_{s}(\tau,t)= \frac{1}{2}\big[F_{B_s}(\mu)]^2e^{-\bar{\Lambda}_s/\tau}
\int\limits_0^\infty \! dk\, e^{-ik t}\phi^{B_s}_+(k,\mu)+\dots\,.
\eea
\begin{figure*}[t]
\begin{center}
\includegraphics[width=0.70\linewidth]{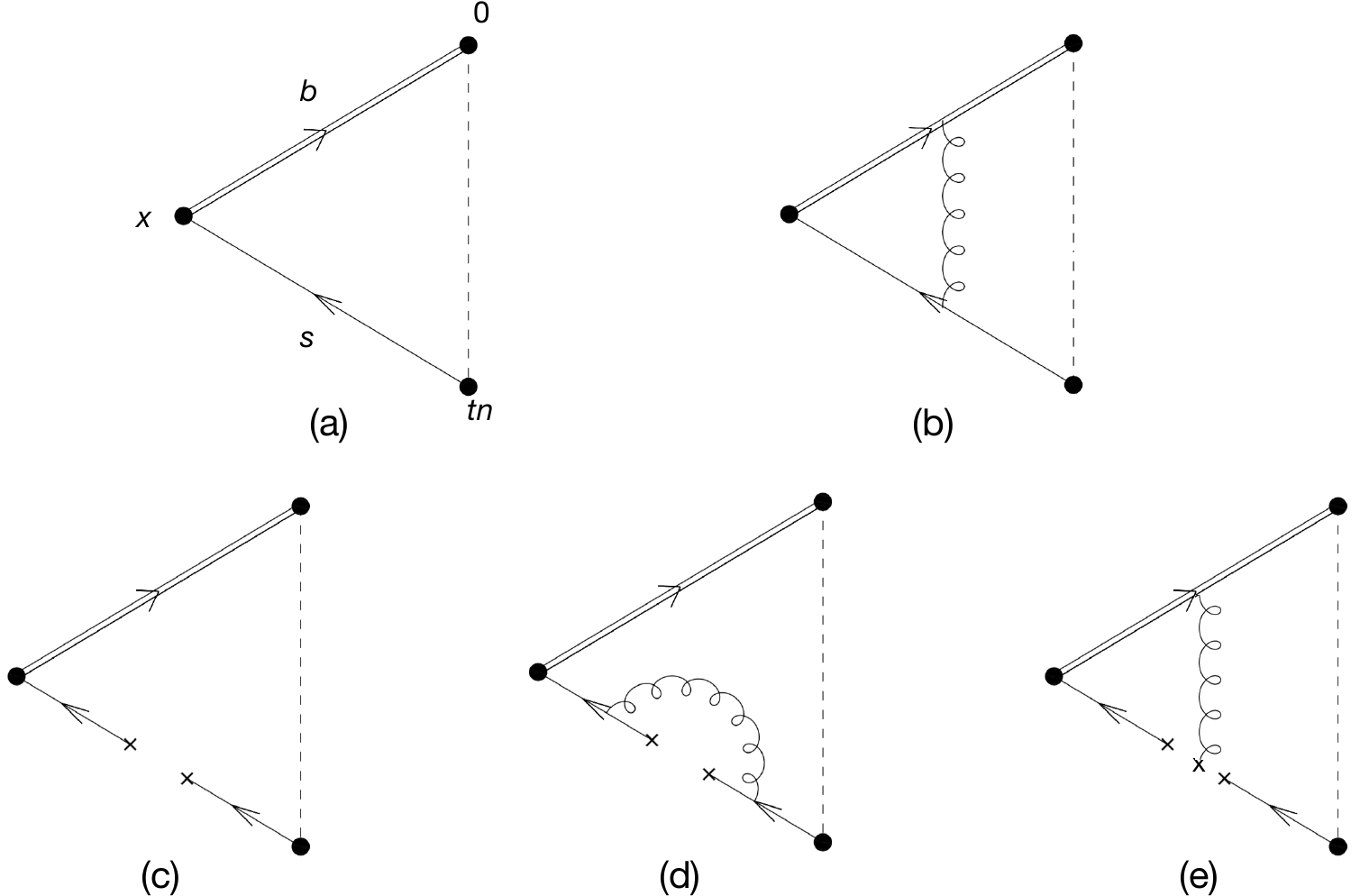}\hskip 20pt 
\caption{The diagrams contributing to the 
correlation function (\ref{eq:BDAsrcorr}): (a)
LO loop; (b) one of the gluon 
radiative 
corrections; (c) quark condensate contribution in LO; (d) one of the gluon radiative corrections to quark condensate; (e) quark-gluon 
condensate contributions. The double line describes the heavy 
($b$-quark) effective field, the point at $x$ corresponds to the 
pseudoscalar interpolating current, the dashed interval connecting the 
points $0$ and $tn$ on the light-cone 
indicates the bilocal operator interpolating the meson DA. All possible diagrams at $\mathcal{O}(\alpha_s)$ can be found in \cite{Braun:2003wx}.
}\label{fig:diags}
\end{center}
\end{figure*}

The next task is to obtain the OPE of the correlation function:
\be
{\cal P}^{\rm OPE}_{s}(\omega,t)={\cal P}^{\rm (pert)}_{s}(\omega,t)+
{\cal P}^{\rm (cond)}_{s}(\omega,t)\,,
\label{eq:OPEampl}
\ee
valid in the region $|\omega|\gg \Lambda_{QCD}$.
The LO perturbative part is described by the diagram in
Fig.~\ref{fig:diags}(a) with a nonzero $s$-quark mass. This diagram is 
a simple loop, where the external momentum transfer takes place 
 through the virtual quark and antiquark lines.
The  $\mathcal{O}(\alpha_s)$ radiative gluon
corrections in ${\cal P}^{\rm(pert)}_{s}$ are exemplified by one of the diagrams shown in 
Fig.~\ref{fig:diags}(b). For these corrections we neglect
the $\mathcal{O}(\alpha_s m_s)$ effects, and use the formulas 
derived in \cite{Braun:2003wx}. The quark condensate 
contribution to ${\cal P}^{\rm(cond)}_{s}$ in LO corresponds to the diagram in
Fig.~\ref{fig:diags} (c) and one of the radiative gluon corrections  
to this term of OPE is shown in Fig.~\ref{fig:diags} (d). In addition, both diagrams in Figs~\ref{fig:diags} (c,e)  contribute to the quark-gluon condensate term. 
Similar to  the sum rule (\ref{eq:SRF}), 
the $SU(3)_{fl}$ violation  reveals itself by the $m_s\neq 0$ and 
$\langle \bar{s}s\rangle\neq \langle \bar{u}u\rangle$ effects, respectively, 
in the perturbative and condensate  parts of the OPE (\ref{eq:OPEampl}).

The perturbative part is
represented in  a form of a dispersion  integral in the
variable $\omega^\prime$:
\bea
\label{eq:pertdisp}
&&{\cal P}^{\rm (pert)}_{s}(\omega,t)=
\frac1{\pi}\int\limits_{m_s}^\infty \!
d\omega'\,\frac{\mbox{Im}\,{\cal P}^{\rm(pert)}_{s}(\omega,t)}{\omega'-\omega}\,.
\eea
Note that the lower limit of the integration is equal to the 
threshold of the quark loop with $m_s\neq 0$ in HQET (cf. the
LO term in Eq.~(\ref{eq:SRF})).
Using the relation (\ref{eq:pertdisp}) in Eq.~(\ref{eq:OPEampl}) and
  performing  Borel transformation, we equate the result to Eq.~(\ref{eq:BDASdisp2}):
\bea
\label{eq:pertdispB}
\frac{1}{2}\big[F_{B_s}(\mu)]^2e^{-\bar{\Lambda}_s/\tau}
\int\limits_0^\infty \! dk\, e^{-ik t}\phi^{B_s}_+(k,\mu)+\dots=
\frac1{\pi}\int\limits_{m_s}^{\infty} 
\!d\omega' \,e^{-\omega'/\tau}\,\mbox{Im}{\cal P}^{\rm(pert)}_{s}
(\omega',t)
+{\cal P}^{\rm(cond)}_{s}(\tau,t)\,.
\eea
Applying the quark-hadron duality approximation, we  equate the
sum of contributions on l.h.s. located above the $B_s$ pole  to the
part of the integral on r.h.s. above the threshold $\omega_{0s}$ which is
taken the same as in the 
sum rule (\ref{eq:SRF}). 
We finally obtain:
\bea
\label{eq:BDASRinit}
\frac{1}{2}\big[F_{B_s}(\mu)]^2e^{-\bar{\Lambda}_s/\tau}
\int\limits_0^\infty \! dk \,e^{-ik t}\phi^{B_s}_+(k,\mu)=
\frac1{\pi}\int\limits_{m_s}^{\omega_{0s}} 
\!d\omega' \,e^{-\omega'/\tau}\,\mbox{Im}{\cal P}^{\rm (pert)}_{s}
(\omega',t)+{\cal P}^{\rm(cond)}_{s}(\tau,t)
\,.
\eea

Calculating the perturbative spectral density and condensate term from the diagrams in Fig.~\ref{fig:diags}, it is possible to reduce both of them 
to the half-Fourier transforms similar to the l.h.s.:
\bea
\label{eq:halfF}
\mbox{Im}{\cal P}^{\rm(pert)}_{s}(\omega',t)=\int\limits_0^\infty \!dk\,
e^{-ikt}\,\mbox{Im}\widetilde{{\cal P}}^{\rm(pert)}_{s}(\omega',k)\,,
~~{\cal P}^{\rm(cond)}_{s}(\tau,t)=\int\limits_0^\infty \!dk\,
e^{-ikt}\,\widetilde{{\cal P}}^{\rm(cond)}_{s}(\tau,k)\,.
\eea
The calculation details are
presented in Appendix. For the 
NLO $\mathcal{O}(\alpha_s)$
corrections, we neglect the $\mathcal{O}(m_s)$ effects and employ the results of \cite{Braun:2003wx} obtained for the correlation function with a massless quark.
The condensate contributions in the form 
(\ref{eq:halfF}) are also inferred from 
the results of \cite{Braun:2003wx} replacing 
$\langle\bar{u}u\rangle\to \langle \bar{s}s\rangle$.

Substituting Eq.~(\ref{eq:halfF}) in Eq.~(\ref{eq:BDASRinit}) and
comparing the integrands on 
both sides, we are in a position to 
read off the $B_s$-meson DA as a function of the variable $k$, in the
same way as it
has been done for the $B$-meson DA in \cite{{Grozin:1996pq},Braun:2003wx} 
(see also \cite{Khodjamirian:2005ea} where analogous sum rules
have been  obtained for the $B$-meson quark-antiquark-gluon DAs ). 
However, as noted already in
  \cite{Grozin:1996pq,Braun:2003wx} and discussed in detail below,  
 the sum rule based on a local OPE  yields a DA which is not 
a smooth function, since the local
condensate contributions produce terms of the form $\phi_+ (k) \sim
\delta(k)$.
The region of small $k$ has to be regularized by introducing a
  nonlocality into the condensate contribution.
In addition, we expect also modifications
of the $B_s$ DA at small $k$, due to the threshold effects from $m_s \neq 0$.
 
In this work, we are eventually interested in
the  inverse moment (\ref{eq:lambBs}). Hence, we will avoid the determination of
the DA shape, noticing that we can directly obtain a sum rule for the inverse moment,  integrating
both sides of \eqref{eq:BDASRinit} over the parameter $t$ for $0 \le t \le \infty$ :
\bea
\label{eq:lambBsSR}
\frac{1}{2}\big[\lambda_{B_s}(\mu)\big]^{-1}\big[F_{B_s}(\mu)]^2e^{-\bar{\Lambda}_s/\tau}=
\frac1{\pi}\int\limits_{m_s}^{\omega_{0s}} 
\!d\omega' \,e^{-\omega'/\tau}
\int\limits_0^\infty \!\frac{dk}{k}\,
\mbox{Im}\widetilde{{\cal P}}^{\rm(pert)}_{s}(\omega',k)
+\int\limits_0^\infty \!\frac{dk}{k}\,
\widetilde{{\cal P}}^{\rm(cond)}_{s}(\tau,k)\,.
\eea
The perturbative part of this sum rule
consists of the LO and NLO parts: 
$$
\mbox{Im}\widetilde{{\cal P}}^{\rm(pert)}_{s}(\omega',k)=
\mbox{Im}\widetilde{{\cal P}}^{\rm(pert,LO)}_{s}(\omega',k)+
\mbox{Im}\widetilde{{\cal P}}^{\rm (pert,NLO)}_{s}(\omega',k)\,.
$$
The LO part, at $m_s\neq 0$, derived in the Appendix reads: 
\bea
\label{eq:lambBsSRLO}
\mbox{Im}\widetilde{{\cal P}}^{\rm (pert,LO)}_{s}(\omega',k)=\frac{
  3}{4\pi}\theta\big(k_{\rm max}(\omega')-k\big)\theta\big(k-
k_{\rm min}(\omega')\big) (k+m_s)\,,
\eea
where
\bea
\label{eq:kminmax}
k_{\rm max,\,min}(\omega')=\omega' \pm \sqrt{\omega^{'2}-m_s^2}\,.
\eea

The NLO part (at $m_s=0$) is taken from \cite{Braun:2003wx};
we only change the  order of integrations. Performing the
$k$-integration of the LO part (\ref{eq:lambBsSRLO}),  
we obtain a more detailed expression for the sum rule (\ref{eq:lambBsSR}):
\bea 
\label{eq:SRlambBs}
&&\big[\lambda_{B_s}(\mu)\big]^{-1}
\big[F_{B_s}(\mu)]^2e^{-\bar{\Lambda}_s/\tau}=
\frac{3}{2\pi^2}\int\limits_{m_s}^{\omega_{0s}} 
\!d\omega' \,e^{-\omega'/\tau}\Bigg[2\sqrt{\omega^{'2}-m_s^2}+
m_s\log\frac{\omega'+\sqrt{\omega^{'2}-m_s^2}}{\omega'-\sqrt{\omega^{'2}-m_s^2}}\Bigg]
\nonumber\\
&&+ \frac{\alpha_s}{\pi^3}\int\limits_0^{\omega_{0s}}\!d\omega'\,e^{-\omega'/\tau}
\Bigg[ \int\limits_{0}^{2\omega'} \!dk\,\widetilde{\rho}_{<}(\omega',k,\mu)
+\int\limits_{2\omega'}^{2\omega_{0s}} \!dk\,\widetilde{\rho}_{>}(\omega',k,\mu)+
\int\limits_{2\omega_{0s}}^{\infty}
\!dk\,\widetilde{\rho}_{>}(\omega',k,\mu)\Bigg]
+C_s(\tau)\,,
\nonumber
\\
\eea
where  the functions $\widetilde{\rho}_{<}$ and $\widetilde{\rho}_{>}$ are presented in
Appendix and 
\bea
C_s(\tau)=2\int\limits_0^\infty \!\frac{dk}{k}\,
\widetilde{{\cal P}}^{\rm(cond)}_{s}(\tau,k)\,,
\label{eq:condtot}
\eea
is the nonperturbative contribution to the sum rule
for inverse moment.  

This contribution is dominated 
by the $s$-quark vacuum condensate, described by the 
diagram in Fig~\ref{fig:diags}(c). As already 
known from \cite{Grozin:1996pq,Braun:2003wx}, 
the local condensate  approximation for this diagram 
yields a divergence. 
The  form  of the local condensate term is easy to obtain,
replacing in the correlation function (\ref{eq:BDAsrcorr}) the vacuum average
of $s$-quark fields by a constant condensate
density, i.e. effectively neglecting the 
momentum flow through the $s$-quark  lines in Fig~\ref{fig:diags}(c). 
The condensate term after Borel transform reads:
\bea
\label{eq:condloc}
{\cal P}^{\rm(cond)}_{s}(\tau,t)=-\frac{1}{2}\langle \bar{s}s\rangle\,,
\eea
yielding 
\bea
\label{eq:condloc2}
\widetilde{{\cal P}}^{\rm(cond)}_{s}(\tau,k)=-\frac{1}{2}\langle \bar{s}s\rangle\delta(k)\,,
\eea
 which, after integration in Eq.(\ref{eq:condtot}), 
indeed results in a divergent contribution to the
 inverse moment. The quark-gluon  and other higher dimension
condensates produce even stronger
singularities. 

The remedy suggested in \cite{Grozin:1996pq,Braun:2003wx} -- which
we also adopt  here -- is to use a nonlocal condensate introduced
earlier \cite{Mikhailov:1986be} in the context of QCD sum rules 
for the pion DA. The quark-antiquark fluctuations in QCD vacuum 
are parameterized as a vacuum expectation value of a bilocal
quark-antiquark operator
\bea
\label{eq:qqbarCorrl}
\langle 0|\bar s(x)[x,0]s(0)|0\rangle =
\langle \bar s s\rangle  \int_{0}^{\infty} \!d\nu \, e^{\nu
  x^2/4}{\mathcal F}(\nu) \,, 
\eea
where the function ${\cal F}(\nu)$ is interpreted as a
distribution of the quark-antiquark vacuum fluctuations with
the virtuality
$\nu$. Expanding the above parameterization at 
small distances, around $x^2=0$, one fixes 
the first two terms of this expansion, matching them to the 
 quark and quark-gluon condensate densities in the local OPE:
\bea
\label{eq:fScond}
\int_0^\infty\!d\nu\, {\cal F}(\nu) =1\,,\qquad
\int_0^\infty\!d\nu\, \nu {\cal F}(\nu) = \frac{m_0^2}{4}\,,
\eea
or, equivalently, 
\bea
{\cal F}(\nu) = \delta(\nu) - \frac{m_0^2}{4} \delta^\prime(\nu) + \ldots\,.
\eea
Here we neglect the possible $O(m_s)$ threshold effects in the 
nonlocal strange-quark condensate and assume that the only 
difference between Eq.~(\ref{eq:qqbarCorrl}) and the nonstrange 
nonlocal condensate is in the quark condensate density. 
The $SU(3)_{fl}$ violation in the first power moment 
in Eq.(\ref{eq:fScond}) amounts to replacing $m_0^2$ by $m_0^2-m_s^2$
(see e.g.  \cite{Bakulev:2002hk}) and is neglected in view of the much larger uncertainty in the parameter $m_0^2$. 
In addition, an exponential fall-off of the nonlocal condensate 
is required at large Euclidean separations $|x^2|\to \infty$.
Following \cite{Braun:2003wx}, 
we choose the two conceivable models, suggested, respectively in
\cite{Mikhailov:1986be} and \cite{Braun:1994jq}:
\begin{align}
\label{eq:Fnu1}
\mbox{model I}:&~~
{\cal F}(\nu) = \delta(\nu-m_0^2/4) \,,
\\
\label{eq:Fnu2}
\mbox{model II}:&~~{\cal F}(\nu) =
\frac{\lambda^{p-2}}{\Gamma(p-2)} \nu^{1-p} e^{-\lambda/\nu}\,,
~~~p= 3+\frac{4\lambda}{m_0^2}\,,
\end{align}
where all above mentioned conditions are satisfied. 
The model II has one free parameter $\lambda$. 
We follow the arguments presented in \cite{Radyushkin:1994xv} (see
also \cite{Bakulev:2002hk}),  where the nonlocal condensate 
is linked  to the
light-quark propagator at large distances and inferred from the 
correlation function of heavy-light currents in HQET. Accordingly, we choose 
this parameter equal to the square of the binding energy in HQET, 
$ \lambda=\bar{\Lambda}^2$ for both $B_s$ and $B$, neglecting the
difference $\bar\Lambda_s- \bar\Lambda$ which is
a second order effect in $m_s$ in Eq.~(\ref{eq:qqbarCorrl}).

Replacing the local condensate density by the nonlocal distribution
(\ref{eq:qqbarCorrl}) leads, 
instead of Eq.~(\ref{eq:condloc}), to  the following  condensate term in the Borel transformed correlation function:
\begin{equation}
\label{eq:Picond}
{\cal P}^{\rm (cond)}_{s}(\tau, t)=-\frac{1}{2}\langle \bar{s}s \rangle 
\int\limits_0^\infty d\nu {\cal F}(\nu)e^{-\frac{\nu}{4\tau^2}-\frac{it\nu}{2\tau}}\,.
\end{equation}
A detailed derivation of this expression is presented in the Appendix.
Equating it to the half-Fourier
representation (\ref{eq:halfF}) and integrating both parts of this equation 
over $0<t<\infty $, we  obtain the condensate term in the sum rule  (\ref{eq:SRlambBs}):
\begin{equation}
\label{Scond}
C_s(\tau)=-2\langle \bar{s}s \rangle \tau
\int\limits_0^\infty \frac{d\nu}{\nu} {\cal F}(\nu)e^{-\frac{\nu}{4\tau^2}}\,.
\end{equation}
Since the nonlocal condensate effectively involves quark and quark-gluon
 condensates, we will not include
the $O(\alpha_s) $ corrections to the local quark condensate, i.e. 
the diagrams  similar to the one in Fig.~\ref{fig:diags}(d) calculated in 
\cite{Braun:2003wx}.
The gluon condensate contribution calculated there and found very 
small is also neglected here. 

\section{Numerical results}
\label{sect:Num}
\begin{table}[!hbt]
\begin{center}
\begin{tabular}{|c|c|c|}
\hline \noalign{\vskip 2pt}
Parameters & Values & Ref. \\ [.7ex]
\hline
\hline
\noalign{\vskip2pt}
Strange quark mass 
&  $\overline{m}_s(2\,\gev)= 93^{+11}_{-5}\,\mev$ &
\cite{Tanabashi:2018oca}   \\[1ex]
\hline 
\noalign{\vskip 1pt}
\multirow{2}{*}{QCD coupling} 
& $\alpha_s(m_Z)= 0.1179 \pm 0.011$ &	
\multirow{2}{*}{\cite{Tanabashi:2018oca,Chetyrkin:2000yt}} \\[0.5ex]
& $\alpha_s \,(1 \,\gev)= 0.458$  
&  \\[1.ex] 
\hline
\noalign{\vskip 2pt}
\multirow{3}{*}{Condensates }
&$ \langle\bar u u \rangle (2\,\gev) =\langle\bar d d \rangle (2\,\gev)= - (288^{+15}_{-13} \,\mev)^3$ 
& 	\multirow{3}{*}{\cite{Tanabashi:2018oca,Ioffe:2002ee}}    \\[0.5ex]
& $\langle\bar s s \rangle/\langle\bar u u \rangle  = 0.8\pm 0.3$ 
&  \\[0.5ex]
& $m_0^2=0.8\pm 0.2\,\gev^2$ & \\[.7ex]
\hline
\noalign{\vskip 1pt}
\multirow{2}{*}{Meson masses}   
&  $m_{B}= (m_{B^\pm}+m_{B^0})/2=(5279.50\pm 0.12)\,\mev$ & 	
\multirow{2}{*}{\cite{Tanabashi:2018oca}}   \\[.5ex] 
& $m_{B_s}= (5366.88\pm 0.17)\,\mev$ & \\[1.ex] 
\hline
\noalign{\vskip 2pt}
HQET binding energy   
&  $\bar{\Lambda}= (0.55\pm 0.06)\,\gev$ & 	\cite{Gambino:2017vkx}   
\\[.7ex]
\hline
\end{tabular}
\caption{Values of the input parameters used in the
    numerical analysis. }
\label{tab:inputs}
\end{center}
\end{table}
We turn to the numerical analysis of the QCD
  sum rule (\ref{eq:SRlambBs}) for the inverse moment $\lambda_{B_s}$.
In parallel, we obtain an estimate for  $\lambda_B$ by putting $m_s=0$
in Eq.~\eqref{eq:SRlambBs} and making the replacements
given in Eq.~\eqref{eq:replacement}. The necessary input
parameters  are listed in Table~\ref{tab:inputs}.

Importantly, instead of the square of the heavy meson decay constant,  we use the  sum rule
(\ref{eq:SRF}) and its $B$-meson counterpart. This has an advantage of canceling out the 
HQET binding energy from the resulting expression for the inverse
moment. Given the relation (\ref{eq:lambdarel}),
we practically only need to specify the parameter $\bar{\Lambda}$ 
in order to  fix the effective
thresholds $\omega_{0}$ and $\omega_{0s}$ from the  differentiated sum
rules and the parameter $\lambda$ in the model II. We use
the most accurate central value obtained from the lattice QCD
simulation of HQET \cite{Gambino:2017vkx,Bazavov:2018omf}\,\footnote{Note that 
$\bar{\Lambda}\simeq m_B-m_b$ is defined in
\cite{Gambino:2017vkx,Bazavov:2018omf},
employing a specific
definition of the $b$  quark mass, adapted to the heavy-quark
expansion of the $B$-meson mass, whereas in the  QCD sum rules,
the $\overline{MS}$ mass of the virtual $b$-quark is conveniently
used.} 
and doubled the uncertainty, to be on a conservative side. Moreover, use of the
sum rule for $F_{B_{(s)}}$ leads to a partial cancellation of the 
renormalization scale and Borel parameter dependences in the sum rule for
$\lambda_{B_{(s)}}$.  

In addition, we have to specify the optimal renormalization scale and
 the interval of Borel parameter. We adopt the default scale 
$\bar{\mu}=3.0$ GeV and the interval $M^2=4.5-6.5$ GeV$^2$ used in the 
numerical analysis of the QCD sum rule for the $B_{(s)}$ decay
constants in full QCD in \cite{Gelhausen:2013wia}.
There one can find a detailed discussion of this choice. We then use the 
rescaling relation (\ref{eq:hqet2}) and obtain the interval
\bea
\label{eq:Borelrange}
\tau=\frac{M^2}{2m_b(\bar{\mu})}=0.5-0.7 ~\gev \,,  
\eea
where the value of the $\overline{MS}$  mass $m_b(\bar{\mu})=4.47$ GeV is obtained by
running from the central value 
$\overline{m}_b(\overline{m}_b)=4.18~\gev$ \cite{Tanabashi:2018oca}.
As a default value we adopt $\tau=0.6\,\gev$. Furthermore, since the
optimal renormalization scale in the sum rule is in the ballpark of the Borel parameter,
it is conceivable to use a not much larger scale $\mu=1.0~\gev$ also in the
HQET sum rule\,\footnote{Obtaining the estimate of the inverse
  moment at $\mu\sim 1\,\gev$, we leave the issue of 
the renormalization scale dependence of the $B_{(s)}$-meson DA \cite{Lange:2003ff} beyond our scope.}.

As a next step, we fix the duality 
thresholds with the procedure 
 described in Sect.~\ref{sec:decayconst}:
\bea
\omega_0 = 1.00\pm 0.12 \,\gev,~~~\omega_{0s}= 1.10\pm 0.13 \,\gev\,.
\label{eq:thresh}
\eea
To assess the $SU(3)_{fl}$ symmetry violation,  we note in passing that 
the ratio of the HQET decay constants calculated from
  the sum rule \eqref{eq:SRF}:
\bea
F_{B_s}(\mu = 1\,\gev)/ F_{B}(\mu = 1\,\gev) = 1.16 \pm 0.08
\eea
is in agreement with the analogous ratio
$f_{B_s}/f_B$  obtained from the lattice QCD
\cite{Aoki:2019cca} and from the sum rules in full QCD 
(see e.g.,\cite{Gelhausen:2013wia}). Hereafter, the errors of our
predictions are estimated incorporating all individual
uncertainties generated by a separate variation of 
each input parameter within its adopted interval. This includes
the parameters listed in Table 1 and the Borel interval (\ref{eq:Borelrange}), whereas the value
of the threshold $\omega_{0(s)}$ is adjusted each time for a given combination of
other inputs.

\begin{table}[h]
\begin{center}
\begin{tabular}{|c|c|c|}
\hline \noalign{\vskip2pt}
Quantity& $\big[\lambda_{B_s}(\mu=1~\gev)\big]^{-1}$& 
$\big[\lambda_{B}(\mu=1~\gev)\big]^{-1}$ \\[1.ex]
\hline \hline \noalign{\vskip2pt}
Perturbative contribution  &$1.34 \pm 0.15 $& $1.17 \pm 0.05$\\[.5ex]
Condensate contribution (model I) &$ 0.66 \pm 0.25 $&$ 1.00\pm 0.24 $\\[.5ex]
Condensate contribution (model II) &$ 1.23 \pm 0.51 $&$ 1.88\pm 0.56$\\[.5ex]
\hline \noalign{\vskip1pt}
total value (model I)&$ 2.00 \pm 0.29 $& $2.17 \pm 0.24$\\[.5ex]
total value (model II)&$ 2.57 \pm 0.53 $& $ 3.05 \pm 0.56 $\\
\hline 
\end{tabular}
\caption{The QCD sum rule prediction for the inverse value of the 
  $B_s$ and $B$ DA inverse moment (in the units GeV$^{-1}$).}
\label{tab:res}
\end{center}
\end{table}

Our numerical results are presented in Table~\ref{tab:res}, where the inverse
values of $\lambda_{B_s}$ and $\lambda_B$ obtained from the sum rule
  (\ref{eq:SRlambBs}) are compared. The
  latter is in the same ballpark as in \cite{Braun:2003wx} (see
  Eq.~(38) there); the difference is caused by
  the deviations of the input parameters, mainly of the quark
  condensate density and $\lambda$.
The condensate contributions are of the same order as the
perturbative ones; note that
  the quark condensate contributions are also enhanced in the correlation
functions with heavy-light currents  in full QCD. 
Here we are mainly interested in the magnitude of the
$SU(3)_{fl}$ symmetry violation. A comparison of separate
contributions to the sum rules 
for $\lambda_{B_s}^{-1}$ and $\lambda_{B}^{-1}$ shows that 
a $\sim 15\%$ decrease in the perturbative part is accompanied by an
up to $\sim 30\%$ increase in the condensate part. However, the
accuracy of the latter estimate
suffers from the large uncertainty
of the ratio of strange and nonstrange condensates. 
We treat the difference between the condensate
contributions obtained with the two models of nonlocal condensate as
an approximate  measure of the accuracy of the 
nonperturbative contributions. Adding this difference to the parametrical
uncertainty in quadrature, we obtain the following intervals for the inverse
moments:
\bea
\lambda_{B_s}= 438 \pm 150 ~\mev\,, ~~ \lambda_B= 383 \pm 153~\mev\,.
\label{eq:mainres}
\eea
The previous result \cite{Braun:2003wx}  $\lambda_B=460\pm 110$
MeV  is in agreement with our estimate. Note that we estimate the
  uncertainties differently and in a more conservative way.
The ratio of the two inverse moments that we predict:
\bea
\frac{\lambda_{B_s}}{\lambda_B}= 1.19 \pm 0.14\,,
\label{eq:ratres}
\eea
is obtained varying in a correlated way all the common inputs
(e.g., the Borel parameter, quark condensate density)
in both the sum rules. Due to the partial cancellations of inputs in this ratio, the resulting uncertainty is
smaller compared  to the individual errors estimated in Eq.~(\ref{eq:mainres}).

The result in Eq.~(\ref{eq:ratres}) can be used in 
future  when a more accurate value
of $\lambda_B$  is available, e.g. from the analysis of 
 the photoleptonic decay $B\to \ell\nu_\ell \gamma$ combined with
   its measurement.

\section{Summary}
\label{sect:Sum}

In this paper we have obtained the first  
estimate of the inverse moment of the leading twist $B_s$-meson DA, assessing the $SU(3)_{fl}$
violation in this important hadronic parameter needed for an accurate 
theoretical description of the $B_s$ exclusive decays.  
We used the HQET sum rule based on the correlation function 
containing  a nonlocal  heavy-light operator 
and a local $B_s$ interpolating current. Instead of aiming at a
determination of the shape of the DA, we obtained a sum rule 
for the inverse moment.
We found that $SU(3)_{fl}$ violation in the inverse moments 
is an appreciable effect, in the same ballpark as for the heavy meson
decay constants. The perturbative contribution to this effect is a
combination of the $\mathcal{O}(m_s)$ term 
computed at LO  and the difference in the quark-hadron duality thresholds. The
latter we fixed with the help of auxiliary  two-point sum rules for
the heavy meson decay constant.  This allows to somewhat reduce
the systematic uncertainty due to the duality approximation.
In the nonperturbative part of the sum rule we employed the nonlocal condensate ansatz which
on one hand effectively includes both quark and quark-gluon
condensates and on the other hand  allows to avoid divergences caused
by the local condensate appearing in
the local OPE.
Using two different model descriptions, we found the condensate
contribution to the $SU(3)_{fl}$ violation, governed in our
approximation by the ratio of strange and nonstrange condensate densities,
to be as important as the perturbative part. 

Our main practical result is 
the ratio of inverse moments of the $B_s$- and $B$-meson DAs 
in which some correlated uncertainties partially cancel. This
ratio indicates that the inverse moment of the $B_s$-meson DA is larger than the one of the $B$-meson, within conservatively estimated uncertainties.
Altogether, the HQET version of QCD sum rules remains an approximate
but the only available tool to investigate the heavy-meson DAs, before the
lattice QCD  methods  become sufficiently developed to   tackle this problem.

\section*{Acknowledgments}
The work of A.K. and Th.M. is supported by the DFG (German Research
Foundation) under grant  396021762-TRR 257 
 ``Particle Physics Phenomenology after the Higgs Discovery''.
The work of R.M. is supported by the Alexander von Humboldt
Foundation through a postdoctoral research fellowship. 
The authors would like to express a special thanks to the Mainz Institute for Theoretical Physics (MITP) of the
DFG Cluster of Excellence PRISMA+ (Project ID 39083149) for its
hospitality and support during the Scientific Program ``Light-Cone
  Distribution Amplitudes in QCD and their Applications''.

\section*{Details of calculation}

\subsection*{Perturbative spectral density at LO}
Here we explain how to compute the LO contribution (the diagram in 
Figure~\ref{fig:diags}(a)) to the 
HQET correlation function  \eqref{eq:BDAsrcorr}. 
Our convention for the light-cone vectors is: 
\begin{equation}
	n^{\mu} \equiv (1,0,0,1), \quad \bar{n}^{\mu} \equiv (1,0,0,-1) \,,
\end{equation}
so that a decomposition of a four-vector $a^\mu$  into its light-cone components 
reads: 
$$a^\mu =\frac{1}{2}\left( a_+ n^\mu+ a_-\bar{ n}^\mu \right) +
a_\perp^\mu$$ 
with $a_+\equiv a.\bar{n} =  \left(a^0+ a^3\right)$ and $a_-\equiv a.n =  \left(a^0- a^3\right)$. 
Adopting the light-cone gauge $(A_+=0)$ removes
the light-like gauge link $[tn,0]\equiv1$. 
  
Contracting the effective heavy-quark and (massive) $s$-quark fields
 into free-field propagators, we obtain for the LO contribution:
\begin{align}
\mathcal{P}_{s}^{\rm (pert,LO)} (\omega, t) =& -3i  \! 
\int d^4x\, e^{-i\omega v\cdot x} 
\nn \\
\times &\int \frac{d^4 \ell}{(2\pi)^4} 
\,e^{-i\ell\cdot(x-tn)} \!\int \frac{d^4 k}{(2\pi)^4} 
e^{i k\cdot x}\, {\rm Tr} \left[
 \frac{\slashed{\ell}+m_s}{\ell^2-m_s^2} 
\gamma_5\slashed{n}  \frac{1+\slashed{v}}{2 k\cdot v} \gamma_5 \right] \nn \\
= & - 3  \!\! \int \frac{d^4 \ell}{(2\pi)^4} 
\,e^{i t(\ell\cdot n)}\! \int \frac{d^4 k}{(2\pi)^4}\!  \int d^4x\, 
e^{-i x\cdot (\omega v+\ell -k)} \frac{4 \ell\!\cdot\! n - 
4 m_s v\!\cdot\! n}{\left(\ell^2-m_s^2\right) \left(2 k\cdot v\right)}\,, 
\end{align}
where the factor $N_c=3$ originates from the colour trace.	
The coordinate integration gives $\delta^{(4)}(k-(\ell+\omega v))$ 
which is removed by the momentum $k$ integration
 and we obtain: 
\bea 
\mathcal{P}_{s}^{\rm (pert,LO)} (\omega, t)  = 3 i \! \int
\frac{d^4 \ell}{(2\pi)^4} \,e^{i t(\ell\cdot n)}
\frac{4 (\ell\cdot n -  m_s v\!\cdot \! n)}{\left(\ell^2-m_s^2\right) 
\left(2(\ell+\omega v)\cdot v\right)}\,,
\eea 
where the  velocity four-vector $v=(1,0,0,0)$.
We need the imaginary part of the above expression 
in the variable $\omega$. To this end, we employ the
Cutkosky rule for both propagators:
\bea 
\frac{1}{p^2 -m^2} \to -2 \pi i \delta (p^2 - m^2) \theta (\pm p_0)\,,  \nn 
\eea 
and get: 
\begin{align}
\label{eq:imexp}
\Im	\mathcal{P}_{s}^{\rm(pert,LO)} (\omega, t) = -3  \! \int &
 \frac{d^4 \ell}{(2\pi)^2} \,e^{i t(\ell.n)}   
\left[  \ell \!\cdot \! n -  m_s  \right] \times \nn \\
&\delta (\ell^2 - m_s^2)\, \delta ( \ell_0 + \omega )\, \theta(-\ell_0-m_s)\, \theta (\ell_0 + \omega)\,. 
	\end{align}
	
Using the adopted convention for the light-cone vectors, 
we replace the four-dimensional integration 
over the $\ell^\mu$ in Eq.~\eqref{eq:imexp} 
with:
$$ 
\int d^4\ell=\int\limits_{-\infty}^{\infty} d\ell_{-}
\int\limits_{-\infty}^{\infty} d\ell_{+}\int d\vec{\ell}_\perp\,.
$$

Integrating over
$d\vec{\ell}_\perp$  together with 
$\delta(\ell^2-m_s^2)= \delta(\ell_+\ell_+ - |\bm{\ell}_\perp|^2 -
m_s^2)$ generates $\pi
\,\theta( \ell_+\ell_- - m_s^2)$. Next, we carry out the 
$d\ell_+$ integration with $\delta(\ell_0 + \omega) = \delta
\left( (\ell_++\ell_-)/2  + \omega \right)$. After that, changing
the variable $\ell_-=-k$ we obtain:
%
\bea
\Im \mathcal{P}_{s}^{\rm (pert,LO)} (\omega, t)& = &
\frac{3}{4\pi} \int\limits_{-\infty}^{+\infty} \!d k
\,e^{-ikt}\,\left( k + m_s \right) \, \theta \left( 2\omega k - k^2 - m_s^2   \right)
\nonumber\\
&=& \frac{3}{4\pi}
 \int\limits_{0}^{\infty} \!d k
\,e^{-ikt}\, \theta\left( k_{\rm max}(\omega)-\omega\right) 
\theta\left( \omega -k_{\rm min}(\omega)\right) \left( k + m_s \right)\,.
\eea
%
where the limits $k_{\rm max} (\omega')$ and $k_{\rm min} (\omega')> 0$
    are defined in  Eq.~(\ref{eq:kminmax}) and in the last equation above
we have used the  quadratic equation with respect to the variable $k$ 
inside the $\theta$ function, reducing the latter to a product
of  the two theta functions. Comparing this equation with the first one in Eq.~(\ref{eq:halfF}) we finally obtain
Eq.~(\ref{eq:lambBsSRLO}).

\subsection*{Perturbative spectral density at NLO}
\label{app:NLO}

Employing the results obtained in \cite{Braun:2003wx},
we use the following functions
determining the NLO
spectral density in the sum rule \eqref{eq:SRlambBs}: 
\begin{align}
\label{eq:rhol}
\widetilde{\rho}_<(\omega ,k,\mu) &= \frac72+\frac{7\pi^2}{24}-
\ln^2\frac{k}{\mu}-\frac52 \ln(x-1)-(x-1)\ln(x-1)
\nonumber\\
&
-\frac12 \ln^2(x-1)-2\ln\frac{k}{\mu}\bigg[1+\ln(x-1)\bigg]+x\ln x+
{\rm Li}_2\left(\frac{1}{1-x}\right)\,,\\
\label{eq:rhog}
\widetilde{\rho}_>(\omega ,k,\mu) &=
-x+\ln(1-x)-2(1-x)\ln(1-x)+2\ln^2(1-x)
+2\ln\frac{k}{\mu}\bigg[x+\ln(1-x)\bigg]\,.
\end{align}
Here ${\rm Li_2}(x)$ is Euler dilogarithm function and $x= 2\omega /k$.

\subsection*{Nonlocal condensate term}

To obtain the nonlocal condensate contribution (\ref{eq:Picond}), we contract the 
$s$-quark fields in the correlation function \eqref{eq:BDAsrcorr} 
into a vacuum average and parametrize it  in
accordance with Eq.~(\ref{eq:qqbarCorrl}): 
\bea 
\langle 0|\bar s_\alpha^i(tn) s_\beta^k(x)|0\rangle =
\langle \bar s s\rangle
\frac{\delta_{\alpha\beta}\delta^{ik}}{3\cdot4}
\int\limits_{0}^{\infty} \!d\nu \, e^{\nu(tn-x)^2/4}{\mathcal F}(\nu) \,.
\label{eq:ssvacnonloc}
\eea
The heavy-quark fields are contracted into a
  HQET propagator. This results in:
\begin{align}
\mathcal{P}_{s}^{(cond)} (\omega, t) =
& \frac{1 }{4} \langle\bar{s}s \rangle \! \int d^4x\, e^{-i \omega v\cdot x}
 \int \limits_0^\infty d\nu\,e^{\nu(tn-x)^2/4}{\mathcal F}(\nu) \!\int
  \frac{d^4 k}{(2\pi)^4} e^{i k\cdot x}\, {\rm Tr} \left[ \gamma_5 \slashed{n} \frac{1+\slashed{v}}{2 k\cdot v} \gamma_5 \right] \nn \\
= & \frac{1}{2}\langle \bar{s}s \rangle  \!\! \int\limits_0^\infty
    d\nu\, {\mathcal F}(\nu) \int \frac{d^4 k}{(2\pi)^4}\!
    \left[\frac{ v\cdot n}{ k\cdot v}\right] 
\int d^4x\, e^{\frac{\nu(tn-x)^2}4-i(\omega v-k)\cdot x} 
\,. 
\end{align}
Redefining the variable $x \to z=x - tn$ so that  $x=z+tn$, and
  using $v\cdot n=1$,  we get 
\bea 
\label{eq:redef}
\mathcal{P}_{s}^{(cond)} (\omega, t) = \frac{1}{2} \langle\bar{s}s\rangle 
\!\! \int \limits_0^\infty \! d\nu\, {\mathcal F}(\nu) e^{-i\omega t} 
\int \frac{d^4 k}{(2\pi)^4} \frac{e^{i t k\cdot n}}{ k\cdot v}\!  
\int\! d^4z\, e^{\frac{\nu z^2}4-i (\omega v-k)\cdot z} \,.
\eea 

The integral over  four-coordinates is obtained 
by completing  the argument of the exponent to a full square, shifting
the integration variables and applying the Wick rotation, $z_0 \to -i z_4$:
\bea
&&\int \!d^4z\, e^{\frac{\nu z^2}{4}-i (\omega v-k)\cdot z} =
\int \!d^4z\, e^{\frac{\nu}{4}\left(z-\frac{2i (\omega
      v-k)}{\nu}\right)^2} 
e^{\frac{(\omega v-k)^2}{\nu}}
\nonumber\\
&&=e^{\frac{(\omega v-k)^2}{\nu}}\int \!d^4z\, e^{\frac{\nu z^2}4} =
-\frac{16i\pi^2}{\nu^2}e^{\frac{(\omega v-k)^2}{\nu}}\,,
\eea
so that
\bea 
\label{eq:cond2}
\mathcal{P}_{s}^{(cond)} (\omega, t) = -i\frac{\langle\bar{s}s\rangle}{2\pi^2} 
\! \int \limits_0^\infty \! \frac{d\nu}{\nu^2}\, {\mathcal F}(\nu) e^{-i\omega t} 
\int \frac{d^4 k} { k\cdot v} 
e^{i t k\cdot n}\, 
e^{\frac{(\omega v-k)^2}{\nu}}.
\eea 
To compute the four-momentum integral, we use the transformation $k\to f=\omega v-k$, and then,
due to $f\cdot n =f_0-f_3$ and $ f \cdot v =f_0$,  
factorize it into three separate integrations:
\bea
\label{eq:int1}
I(\omega,t)\equiv \int \frac{d^4 k}{ k\cdot v} e^{i t k\cdot n}\, 
e^{\frac{(\omega v-k)^2}{\nu}}= e^{i\omega t} \int \frac{d^4f}{(\omega v-f)\cdot
  v}\, e^{-i t(f\cdot n)+\frac{f^2}{\nu}}
\nonumber\\
=e^{i\omega t}\int\limits_{-\infty}^{+\infty}\frac{df_0}{\omega-f_0} \,
e^{-i tf_0+\frac{f_0^2}{\nu}}
\int\limits_{-\infty}^{+\infty} df_3 \,e^{i t f_3-\frac{f_3^2}{\nu}}\,
2\pi\int\limits_0^\infty d|\vec{f}_\perp||\vec{f}_\perp| \,e^{-\frac{|\vec{f}_\perp|^2}{\nu}}\,,
\eea
where we also used that $v^2=1$.  The integral  over the two-dimensional plane $f_{1,2}$ taken 
in the polar  coordinates
with $|\vec{f}_\perp|=\sqrt{f_1^2+f_2^2}$  is equal to 
$\pi\nu$.  Completing the arguments of
exponential functions in the integrals over the $f_0$, $f_3$, we
integrate
over $f_3$ and, after the shift of the variable $$f_0\to
\tilde{f}_0=f_0-it\nu/2\,,$$
we get
\bea
\label{eq:int2}
I(\omega,t)&=&\pi\nu e^{i\omega t}\!\int\limits_{-\infty}^{+\infty}\!\frac{df_0}{\omega-f_0} \,
e^{\frac{(f_0-it\nu/2)^2}{\nu}} e^{\frac{t^2\nu}{4}}
\!\int\limits_{-\infty}^{+\infty}\! df_3
e^{-\frac{(f_3-it\nu/2)^2}{\nu}} e^{-\frac{t^2\nu}{4}}
\nonumber\\
&=&(\pi\nu)^{3/2}e^{i\omega t}\!\int\limits_{-\infty}^{+\infty}\!\frac{df_0}{\omega-f_0} \,
e^{\frac{(f_0-it\nu/2)^2}{\nu}}=
-(\pi\nu)^{3/2}e^{i\omega t}
\!\int\limits_{-\infty}^{+\infty}\!\frac{d\tilde{f}_0}{\tilde{f}_0+
\frac{it\nu}2-\omega} \,
e^{\frac{\tilde{f}_0^2}{\nu}}\,.
\eea
Substituting this expression in Eq.~(\ref{eq:cond2}), we obtain
\bea 
\label{eq:cond3}
\mathcal{P}_{s}^{(cond)} (\omega, t) = i\frac{\langle\bar{s}s\rangle}{2\sqrt{\pi}} 
\! \int \limits_0^\infty \! \frac{d\nu}{\sqrt{\nu}}\, {\mathcal F}(\nu) 
\!\int\limits_{-\infty}^{+\infty}\!\frac{d\tilde{f}_0}{\tilde{f}_0+
\frac{it\nu}2-\omega} \,
e^{\frac{\tilde{f}_0^2}{\nu}}.
\eea 
At this stage it is convenient to perform the Borel transformation:
\bea 
\label{eq:cond4}
\mathcal{P}_{s}^{(cond)} (\tau, t) = i\frac{\langle\bar{s}s\rangle}{2\sqrt{\pi}} 
\! \int \limits_0^\infty \! \frac{d\nu}{\sqrt{\nu}}\, {\mathcal F}(\nu) 
\!\int\limits_{-\infty}^{+\infty}\!d\tilde{f}_0 \, 
e^{\frac{\tilde{f}_0^2}{\nu}} e^{-\frac{\tilde{f}_0+it\nu/2}{\tau} \,}.
\eea 
Applying the  Wick rotation $\tilde{f}_0\to if_4$  we integrate:
\bea 
\label{eq:intf4}
&&\int\limits_{-\infty}^{+\infty}\!d\tilde{f}_0 \, 
e^{\frac{\tilde{f}_0^2}{\nu}} e^{-\frac{\tilde{f}_0+it\nu/2}{\tau} \,}
=i \!\int\limits_{-\infty}^{+\infty}\!d f_4 \, 
e^{\frac{-f_4^2}{\nu}} e^{-i\frac{f_4+t\nu/2}{\tau} \,}
\nonumber\\
&&=i \! \int\limits_{-\infty}^{+\infty}\!d f_4 \, 
e^{-\frac{(f_4+i\nu/(2\tau))^2}{\nu}} e^{-\frac{\nu}{4\tau^2}}
e^{-\frac{i t\nu}{2\tau}}
=
i \sqrt{\pi}\sqrt{\nu} e^{-\frac{\nu}{4\tau^2}} e^{-\frac{i t\nu}{2\tau}}\,,
\eea 
and, using the above result in Eq.~(\ref{eq:cond4}),
finally reproduce Eq.~(\ref{eq:Picond}).


\begin{thebibliography}{99}	

\bibitem{Grozin:1996pq}
A.~Grozin and M.~Neubert,
Phys.\ Rev.\ D \textbf{55} (1997), 272-290\,.

\bibitem{Szczepaniak:1990dt}
A.~Szczepaniak, E.~M.~Henley and S.~J.~Brodsky,
Phys.\ Lett.\ B \textbf{243} (1990), 287-292\,.

\bibitem{Beneke:2000wa}
M.~Beneke and T.~Feldmann,
Nucl.\ Phys.\ B \textbf{592} (2001), 3-34\,.

\bibitem{Beneke:1999br}
M.~Beneke, G.~Buchalla, M.~Neubert and C.~T.~Sachrajda, 
Phys.\ Rev.\ Lett.\  \textbf{83} (1999), 1914-1917\,. 


\bibitem{Korchemsky:1999qb}
G.~P.~Korchemsky, D.~Pirjol and T.~Yan,
Phys.\ Rev.\ D \textbf{61} (2000), 114510\,.

\bibitem{DescotesGenon:2002mw}
S.~Descotes-Genon and C.~Sachrajda,
Nucl.\ Phys.\ B \textbf{650} (2003), 356-390\,.


\bibitem{Bosch:2003fc}
S.~Bosch, R.~Hill, B.~Lange and M.~Neubert,
Phys.\ Rev.\ D \textbf{67} (2003), 094014\,.


\bibitem{Khodjamirian:2005ea}
A.~Khodjamirian, T.~Mannel and N.~Offen,
Phys.\ Lett.\ B \textbf{620} (2005), 52-60;
Phys.\ Rev.\ D \textbf{75} (2007), 054013\,.


\bibitem{Aoki:2019cca}
S.~Aoki \textit{et al.} [Flavour Lattice Averaging Group],
Eur.\ Phys.\ J.\ C \textbf{80} (2020) no.2, 113.

\bibitem{Jamin:2001fw}
M.~Jamin and B.~O.~Lange,
Phys.\ Rev.\ D \textbf{65} (2002), 056005.

\bibitem{Gelhausen:2013wia}
P.~Gelhausen, A.~Khodjamirian, A.~A.~Pivovarov and D.~Rosenthal,
Phys.\ Rev.\ D \textbf{88} (2013), 014015;
Erratum: [Phys.\ Rev.\ D {\bf 89}, 099901 (2014)], [Phys.\ Rev.\ D {\bf 91}, 099901 (2015)].

\bibitem{Kane:2019jtj}
C.~Kane, C.~Lehner, S.~Meinel and A.~Soni, 
[arXiv:1907.00279 [hep-lat]]. 

\bibitem{Desiderio:2020oej}
A.~Desiderio, R.~Frezzotti, M.~Garofalo, D.~Giusti, M.~Hansen, V.~Lubicz, G.~Martinelli, C.~T.~Sachrajda, F.~Sanfilippo, S.~Simula and N.~Tantalo, 
[arXiv:2006.05358 [hep-lat]]. 

\bibitem{Beneke:2003zv}
M.~Beneke and M.~Neubert, 
Nucl.\ Phys.\ B \textbf{675} (2003), 333-415. 

\bibitem{Beneke:2011nf}
M.~Beneke and J.~Rohrwild,
Eur.\ Phys.\ J.\ C \textbf{71} (2011), 1818.


\bibitem{Beneke:2018wjp}
M.~Beneke, V.~Braun, Y.~Ji and Y.~Wei,
JHEP \textbf{07} (2018), 154.

\bibitem{Wang:2018wfj}
Y.M.~Wang and Y.~Shen,
JHEP \textbf{05} (2018), 184.

\bibitem{Broadhurst:1991fc}
D.~J.~Broadhurst and A.~G.~Grozin,
Phys. Lett. B \textbf{274}, 421-427 (1992).

\bibitem{Braun:2003wx}
V.~Braun, D.~Ivanov and G.~Korchemsky,
Phys.\ Rev.\ D \textbf{69} (2004), 034014.

\bibitem{Bagan:1991sg}
E.~Bagan, P.~Ball, V.~M.~Braun and H.~G.~Dosch,
Phys.\ Lett.\ B \textbf{278} (1992), 457-464.

\bibitem{Neubert:1991sp}
M.~Neubert,
Phys.\ Rev.\ D \textbf{45} (1992), 2451-2466.

\bibitem{Shuryak:1981fza}
E.~V.~Shuryak,
Nucl. Phys. B \textbf{198} (1982), 83-101\,.

\bibitem{Grozin:2005iz}
A.~G.~Grozin,
Int. J. Mod. Phys. A \textbf{20}, 7451-7484 (2005).

\bibitem{Mikhailov:1986be}
S.~Mikhailov and A.~Radyushkin,
JETP Lett. \textbf{43} (1986), 712;
%
Phys. Rev. D \textbf{45} (1992), 1754-1759\,.

\bibitem{Bakulev:2002hk} 
A.~P.~Bakulev and S.~V.~Mikhailov,
Phys.\ Rev.\ D {\bf 65}, 114511 (2002).



\bibitem{Braun:1994jq}
V.~Braun, P.~Gornicki and L.~Mankiewicz,
Phys. Rev. D \textbf{51} (1995), 6036-6051.

\bibitem{Radyushkin:1994xv}
A.~V.~Radyushkin,
[arXiv:hep-ph/9406237 [hep-ph]],
In *Minneapolis 1994, Proceedings, Continuous advances in QCD* 238-248\,.




\bibitem{Gambino:2017vkx}
P.~Gambino, A.~Melis and S.~Simula,
Phys. Rev. D \textbf{96} (2017) no.1, 014511\,.


\bibitem{Bazavov:2018omf}
A.~Bazavov \textit{et al.} [Fermilab Lattice, MILC and TUMQCD],
Phys. Rev. D \textbf{98}, no.5, 054517 (2018).

\bibitem{Tanabashi:2018oca}
M.~Tanabashi \textit{et al.} [Particle Data Group],
Phys. Rev. D \textbf{98}, no.3, 030001 (2018)\,.

\bibitem{Chetyrkin:2000yt}
K.~Chetyrkin, J.~H.~K\"uhn and M.~Steinhauser,
Comput. Phys. Commun. \textbf{133}, 43-65 (2000)\,.


\bibitem{Ioffe:2002ee}
B.~Ioffe,
Phys. Atom. Nucl. \textbf{66}, 30-43 (2003).

\bibitem{Lange:2003ff}
B.~O.~Lange and M.~Neubert,
Phys. Rev. Lett. \textbf{91}, 102001 (2003).

\end{thebibliography}
\end{document}